\documentclass[useAMS]{mn2e}
\usepackage{psfig}
\newcommand{\be}{\begin{equation}}
\newcommand{\en}{\end{equation}}
	
\def\zabs{$z_{\rm abs}$}

\def\lya{Ly$\alpha$ }

\def\h2{H$_2$}

\def\kms{km~s$^{-1}$}
\begin{document}
%
%\maketitle
\title[sub-DLA at \zabs = 1.36]{
Multiwavelength investigation of a near-solar metallicity sub-DLA
at $z_{\rm abs}$~=~1.3647 towards 
PKS 0237-233\thanks{Based on observations carried out at the 
European Southern Observatory  with UVES in the course of the Large Program 
"The Cosmic Evolution of the IGM" 166.A-0106 
on the 8.2~m VLT telescopes Kuyen operated at Paranal Observatory; Chile.}  
}
\author[R. Srianand, N. Gupta \& P. Petitjean]{Raghunathan Srianand$^{1}$, Neeraj Gupta$^{2}$,  Patrick Petitjean$^{3,4}$\\
$^1$IUCAA, Post Bag 4, Ganeshkhind, Pune 411 007, India\\
~email: anand@iucaa.ernet.in\\ 
$^2$NCRA, Post Bag 3,  Ganeshkhind, Pune 411 007, India\\
~email:neeraj@ncra.tifr.res.in\\
$^3$Institut d'Astrophysique de Paris -- Universit\'e Pierre et Marie Curie, 98bis Boulevard 
Arago, F-75014 Paris, France\\
$^4$LERMA, Observatoire de Paris, 61 Rue de l'Observatoire,
   F-75014 Paris, France\\
~email:petitjean@iap.fr\\}
\date{Received xxx / Accepted xxx}
%\offprints{R. Srianand}
\maketitle
\begin{abstract}
We searched for 21-cm absorption associated with the \zabs = 1.3647 absorption
system toward the gigahertz peaked-spectrum source PKS~0237$-$233 using the
Giant Metrewave Radio Telescope. A high quality UVES spectrum shows that
C~{\sc i} and C~{\sc i$^*$} are detected at this redshift together with C~{\sc ii$^*$}, Mg~{\sc i}, 
Mg~{\sc ii}, Si~{\sc ii}, Al~{\sc ii}, Fe~{\sc ii} and Mn~{\sc ii}.
The complex profiles, spread over $\sim 300$ \kms, are fitted with 21 Voigt profile components.
% at \zabs = 1.3647 in the UVES spectrum of the gigahertz 
%peaked-spectrum source PKS~0237$-$233.  
None of these components are detected in 
21-cm absorption down to a detection limit of $\tau(3\sigma)\le3\times10^{-3}$ 
(or $N$(H~{\sc i})/$T_{\rm S} \le10^{17}$~cm$^{-2}$~K$^{-1}$).
% in a radio spectrum obtained with the Giant Metrewave Radio Telescope. 
% that show H$_2$ absorption, where the gas is cold.
We derive log~[$N$(H~{\sc i}) cm$^{-2}$]$\le$19.30$\pm$0.30 
using the \lya absorption line detected in the IUE spectrum of the quasar. 
%From the measured total column density of O~{\sc i} we infer that 
Mg~{\sc ii}, Si~{\sc ii} and Al~{\sc ii} column densities are consistent 
with near solar metallicity and we measure [O/H] $\ge-0.33$. 
%measured column densities of Mg~{\sc ii}, 
%Si~{\sc ii} and Al~{\sc ii} are consistent with near solar metallicity.
%
Using photoionization models constrained by the fine-structure excitations
of C~{\sc i} and C~{\sc ii}, and the 21-cm optical depth, we show that the 
C~{\sc i} absorption arises predominantly either in a warm 
ionized medium (WIM) or  warm neutral medium (WNM)
in ionization and
thermal equilibrium with the meta-galactic UV background dominated by QSOs 
and star forming galaxies. 
The estimated thermal pressure of the gas is of the same order of
magnitude over different velocity ranges through the absorption profile
($2.6\le {\rm log [}P/k {\rm (cm^{-3} K)]}\le 4.0$). 
The gas-phase metallicity corrected for ionization is Z $\ge 0.5$ Z$_\odot$ 
with a signature of iron co-production elements being under abundant 
compared to $\alpha-$process elements by $\sim$0.5 dex.  
At \zabs$\ge$1.9, C~{\sc i} absorption is usually associated with H$_2$ absorption 
arising from cold gas in damped Lyman-$\alpha$ systems.
This system and the $z_{\rm abs}$~=~2.139 toward
Tol~1037$-$270 are the only two systems known which show 
that C~{\sc i} absorption can also be detected in warm gas provided 
the metallicity is high enough.
Interestingly, both the systems are  part of unusual concentrations of absorption lines.
\end{abstract}

\begin{keywords}
Quasars: absorption lines --
Quasars: individual: PKS~0237$-$233             
\end{keywords}
\section{Introduction}

The diffuse gas in the interstellar medium (ISM) is known to exhibit
a wide range of physical conditions (temperature, density, radiation
field and chemical composition). 
These physical quantities are influenced by the local star formation, 
the cosmic ray energy density, photoelectric heating by dust as well 
as mechanical energy input from both impulsive disturbances such 
as supernova explosions (McKee \& Ostriker, 1977)
as well as more steady injection of energy in the form of winds
from stars (Abbott, 1982). Understanding these physical conditions
and the processes that maintain these conditions are important for 
our understanding of galaxies and their evolution. 

The relative populations of the ground and excited levels of C~{\sc i} 
have been used to discuss the particle density, the ambient UV 
radiation field and the temperature of the cosmic-microwave background 
radiation  (see Bahcall, Joss \& Lynds 1973; Meyer et al., 1986; 
Songaila et al. 1994; Ge, Bechtold \& Black, 1997; Roth \& Bauer, 1999; 
Srianand et al. 2000; Silva \& Viegas, 2002). In the Galactic ISM,  
fine-structure excitation of C~{\sc i} has been used to study the 
distribution of thermal pressure (see Jenkins \& Tripp 2001). At high 
redshifts C~{\sc i} absorption is very rare and detected in only a handful
of systems ($\simeq 15\%$ of the damped Lyman $\alpha$ systems 
(DLAs) at \zabs $\ge1.9$). Usually, 
C~{\sc i} absorption in these systems is closely associated with 
low temperature, high density gas that often shows associated H$_2$ 
absorption  (Ge \& Bechtold, 1999; Petitjean et al. 2000; Ledoux, Petitjean \& 
Srianand 2003; Srianand et al. 2005). The only exception is the 
\zabs = 2.139 system
%2.59471 and 2.59486 components 
towards Tol 1037$-$270 (Srianand \& Petitjean, 2001) that originates from a solar 
metallicity sub-DLA 
(i.e. systems that show damping wings with log N(H~{\sc i})$<$ 20.3) 
very much similar to the one we shall discuss here in which
the absence of detectable H$_2$ absorption lines suggests that 
the gas temperature is probably high.
% in these two components.

Like C~{\sc i}, C~{\sc ii$^*$} absorption has been detected in every system
where H$_2$ is present (Srianand et al. 2005). However, C~{\sc ii$^*$} 
has also been seen in a considerable fraction of DLAs ($\simeq 50\%$) 
without H$_2$ (Wolfe, Prochaska \& Gawiser 2003; Wolfe, Gawiser \& Prochaska 2003; 
Srianand et al. 2005). It has been
found that C~{\sc ii$^*$} absorption is always detected when 
log~$N$(H~{\sc i})~$\ge$~21 and also in systems with high metallicity 
irrespective of $N$(H~{\sc i}). As C~{\sc ii}$^*$ 
is an important coolant of the ISM, the corresponding C~{\sc ii$^*$} column 
density can be used to discuss the cooling rate in the absorbing gas
(Wolfe et al. 2004).

The nature and physical conditions in the H~{\sc i} gas can also be
probed by 21-cm absorption lines. The searches for 21-cm absorption in 
DLAs at \zabs$\ge2$ have mostly resulted in null detections with 
typical spin temperatures $\ge10^3$ K (see Table 3 of Kanekar \& Chengalur 
2003 and Table 1 of Curran et al. 2005). 
Based on the lack of 21-cm and H$_2$ absorption in most of the systems it has been
inferred that most of DLAs actually trace a warm neutral phase
(see also Petitjean, Srianand \& Ledoux 2000).
However one of the possible uncertainties to interpret the 21~cm 
data could be the unknown covering factor of the H~{\sc i} gas in front
of the radio source (see Briggs \& Wolfe 1983 for a discussion).

Despite obvious advantages of studying all the above indicators
simultaneously, such studies are possible only for a few systems
[at \zabs = 2.8110 towards PKS~0528$-$250 (Carilli et al., 1996;
Srianand \& Petitjean 1998), \zabs = 1.944 towards Q1157$+$014 
(Wolfe \&  Briggs 1981; Srianand et al. 2005) and  \zabs = 2.03954 towards 
PKS 0458$-$020 (Wolfe et al. 1985; Briggs et al 1989; Heinm\"uller et al. 
2005)]. This is mainly because only a small fraction of the QSOs
with interesting DLAs are radio-loud and, the redshifted 21-cm frequency range
in many cases is not covered by the present-day radio telescopes or is
affected by strong radio frequency interferences.

The UVES  spectrum of the compact source PKS~0237$-$233,
that was obtained as part of  the ESO-VLT large programme
``The Cosmic Evolution of the IGM'' (Bergeron et al. 2004),
shows C~{\sc i}, C~{\sc i$^*$} and C~{\sc ii$^*$} absorption
lines spread over $\sim$300 \kms~ in a strong Mg~{\sc ii} 
sub-DLA system at \zabs = 1.365.
The wavelength range into which the corresponding 21-cm absorption 
is redshifted can be covered by the Giant Metrewave Radio Telescope (GMRT)
 610 MHz band.
Here, we present the analysis of this system  using the high-resolution
optical spectrum obtained with UVES/VLT, a low-resolution UV spectra
obtained with IUE and the GMRT radio spectrum covering the redshifted
21-cm absorption.

\section{Observations and data reduction}

\begin{figure*}
\begin{center}
\psfig{figure=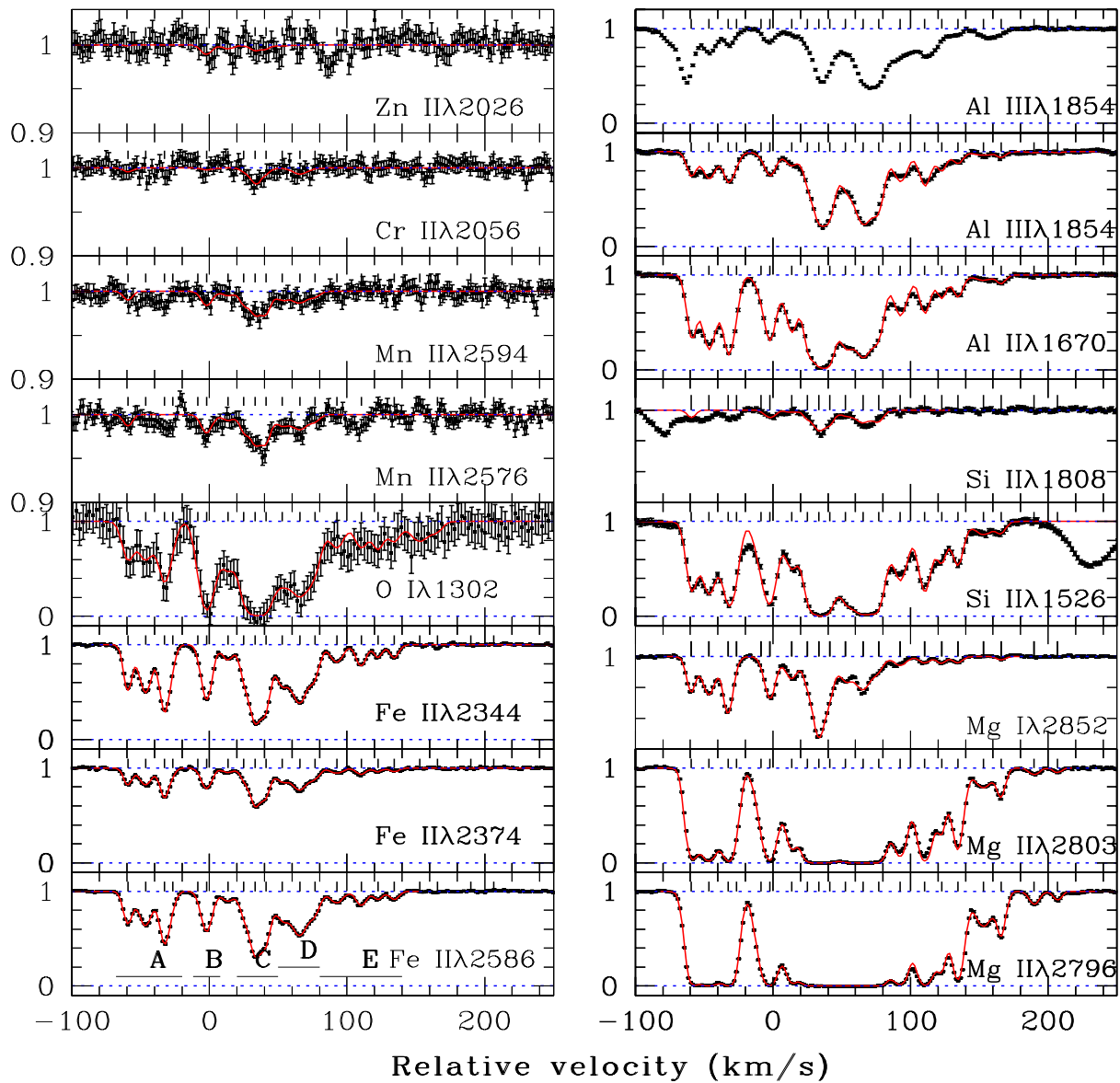,height=17cm,width=18.0cm,angle=0}
\caption[]{
Velocity plots of the absorption line system 
centered at \zabs~=~1.3647. The points with error bars are the 
observations and the continuous curve shows the Voigt profile
fit to the absorption profiles. Tick marks in each panel 
indicate the location of individual components. 
We identify 21 distinct components
based on the fits to Mg~{\sc ii} and Fe~{\sc ii} profiles. 
The distinct velocity ranges discussed in detail in the text
are indicated in the bottom left panel.
}
\label{fig1}
\end{center}
\end{figure*}
\subsection{Optical UVES observations}

The optical spectrum of PKS 0237$-$233 used in this study was obtained
using the Ultraviolet and Visible Echelle Spectrograph 
(UVES; Dekker et al. 2000) mounted on the ESO Kueyen 8.2~m telescope
at the Paranal observatory in the course of the ESO-VLT large programme 
"The Cosmic Evolution of the IGM''. PKS 0237$-$233 was observed through 
a 1$^{\prime\prime}$ slit (with a seeing typically $\leq $0.8 $^{\prime\prime}$)
for $\sim$12 hr with central wavelengths adjusted to 
3460\AA~ and 5800\AA~ in the blue and red arms, respectively, using dichroic no. 1 
and for another $\sim$14 hr with central wavelengths at 4370\AA~ and 8600\AA~ 
in the blue and red arms with dichroic no. 2.
The raw data were reduced using the latest version of the UVES pipeline 
(Ballester et al. 2000) which is available as a dedicated context of
the MIDAS data reduction software. The main function of the pipeline 
is to perform a precise inter-order background subtraction for science 
frames and master flat-fields, and to allow an optimal extraction of 
the object signal, rejecting cosmic ray impacts and performing 
sky-subtraction at the same time.  The reduction is checked step by step. 
Wavelengths are corrected to vacuum-heliocentric values and individual 
1D spectra are combined. Air-vacuum conversions and heliocentric 
corrections were done using standard conversion equations 
(Edl\`en 1966; Stumpff 1980). Addition of individual exposures is 
performed using a sliding window  and weighting the signal by the errors 
in each pixel. Great care was taken in computing the error spectrum while 
combining the individual exposures. Our final error is the quadratic sum 
of appropriately interpolated weighted mean errors and possible errors due 
to scatter in the individual flux measurements.  
The final combined spectrum covers the wavelength range of 3000-10,000\AA. A typical 
S/N $\sim$ 60-80 per pixel (of 0.035~\AA) is achieved in 
the whole wavelength range of interest. The detailed quantitative 
description of data calibration 
is presented in Aracil et al. (2004) and Chand et al. 
(2004).

\subsection{Radio Observations}

The radio spectrum 
%to investigate the 21~cm absorption 
%originating from the \zabs=1.365 system towards PKS 0237$-233$ 
was obtained on 2004, Feb 9 using  GMRT.  
The local oscillator chain was tuned to centre the baseband at the 
corresponding redshifted 21-cm frequency.  The FX correlator system 
at GMRT splits the baseband into 128 channels, yielding a velocity 
resolution of 3.9 \kms~ for our chosen baseband bandwidth of 1 MHz.
We observed standard flux calibrators 3C48 or 3C147 every two hours 
to correct for the amplitudes and bandpass variations.  Since the background 
radio source, PKS 0237$-$233 is unresolved for the GMRT array at the 
observing frequency, observing a separate phase calibrator was not 
required.  A total 8 hrs of on-source data were acquired in both the 
circular polarization channels RR and LL. 
%%%%

The radio data were reduced in the standard way using the Astronomical 
Image Processing System (NRAO AIPS package). After the initial flagging 
and calibration,  source and calibrator data were examined to flag
and exclude the baselines and timestamps affected by Radio Frequency 
Interference (RFI).  
It was noticed that data on 3C~147 was not 
particularly good as compared to 3C~48.  Of the 4 time-scans (exposures) 
on 3C~48 and 3C~147, 3C147 contributed only to the last one.  
So we used only 3C~48 for the bandpass calibration.  Applying these 
complex gains and bandpass, a continuum map of the source was made 
using line free channels.
 Using this map as a model, self-calibration 
complex gains were determined and applied to all the frequency channels.  
The same continuum map was then used to subtract the continuum emission
(a flux density of 5.1 Jy).  
The continuum subtracted data set was then imaged separately in 
stokes RR and LL to get three-dimensional (with third axis as frequency) 
data cubes.  Spectra at the quasar position were extracted from these 
cubes and compared for consistency.  The two polarization channels were 
then combined to derive the final stokes I spectrum, which was then shifted 
to heliocentric frame.  

\subsection{IUE spectrum}

An IUE spectrum of PKS 0237$-$233 was obtained by Richard F. Green
as part of an observing programme to study high-redshift
quasars (IUE Program ID: QSDRG and IUE image number: LWR10935).
The spectrum was obtained with the large aperture (approximately 
10 x 20 arc sec) and an exposure time of 25,800~s. The typical
spectral resolution is $\sim$ 6 \AA~ and the signal to noise per pixel
$\sim5$. We used the pipeline calibrated IUE spectra obtained directly from  
Multimission Archive at Space Telescope (MAST). 

%%%%%%%%%%%%%%%%%%%%%%%%%%%%%%%%%%%%%%%%%%%%%%%%%%%
\section{\zabs = 1.3647 system}
\subsection{Neutral and singly ionized species}
\begin{figure}
\begin{center}
\psfig{figure=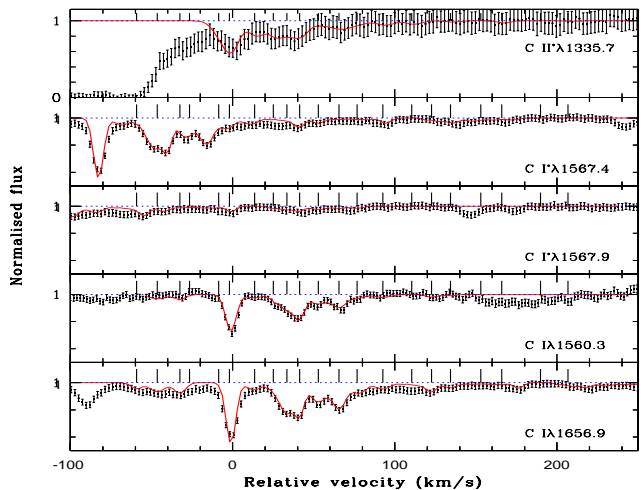,height=7cm,width=9.0cm,angle=0}
\caption[]{Excited fine-structure lines of C~{\sc i} and C~{\sc ii}
plotted on a velocity scale with respect to \zabs = 1.3647. Points with
errorbars are the observed data and the best fit is shown
as a continuous curve. The tick marks indicate the location of distinct 
Voigt profile components.
}
\label{fig2}
\end{center}
\end{figure}
\begin{table*}
\begin{center}
\caption{Voigt profile fitting results for individual components:$^1$}
{\scriptsize
\begin{tabular}{lcccccccccc}
\hline
\multicolumn{1}{c}{$z_{\rm abs}$}&$b$(km/s)& $N$(Mg~{\sc i})& $N$(Mg~{\sc ii})& $N$(O~{\sc i})&$N$(Si~{\sc ii})&$N$(Fe~{\sc ii})&$N$(C~{\sc ii}$^*$) & $N$(C~{\sc i}) & $N$(C~{\sc i}$^*$)&$N$(H I)$^2$\\ % {${\rm N(H I)*100 K \over T_S}$}\\
\hline
\hline
\multicolumn{11}{c}{Velocity range A}\\
1.364232& 2.6(0.2)& 11.21(0.01)& 13.97(0.01)& 13.58(0.08)& 13.58(0.02)&12.89(0.01)& $b$     & 11.80(0.19)&   $\le12.00$&$\le$18.36 \\
1.364334& 5.1(0.2)& 11.38(0.01)& 13.47(0.01)& 13.62(0.06)& 13.56(0.01)&13.04(0.01)& $b$     & 12.12(0.10)&   $\le12.00$&$\le$18.66 \\
1.364443& 3.0(0.2)& 11.50(0.01)& 13.73(0.02)& 13.93(0.09)& 13.77(0.02)&13.21(0.01)& $b$     & 12.07(0.12)&   $\le12.00$&$\le$18.43 \\
1.364490& 2.1(0.7)& 10.00(0.10)& 12.56(0.01)& 12.85(0.31)& 13.04(0.02)&12.21(0.06)& $b$     & $\le11.70$ &   $\le12.00$&$\le$18.30 \\
\multicolumn{2}{c}{Total}&11.86(0.01)&14.26(0.01)&14.23(0.05)&14.16(0.01)&13.56(0.01) & $b$    & 12.49(0.07)&   $\le12.30$&$\le$19.21\\
\\
\multicolumn{11}{c}{Velocity range B}\\
1.364632& 5.1(0.1)&  9.32(0.65)& 12.46(0.01)&  $\le 13.0$& 13.03(0.02)& $\le11.50$ & 12.82(0.06)& $\le11.70$    &   $\le12.00$ &$\le$18.66\\
1.364685& 4.0(0.1)& 11.37(0.01)& 13.46(0.01)& 14.80(0.16)& 13.69(0.01)&13.07(0.01) & 13.13(0.04)& 12.92(0.01)& 12.26(0.02)     &$\le$18.56\\
\multicolumn{2}{c}{Total}&11.37(0.01)&13.50(0.01)&14.80(0.16)&13.77(0.01)&13.07(0.01)& 13.30(0.03)& 12.92(0.01)& 12.26(0.02)&$\le$19.07\\
\\
\multicolumn{11}{c}{Velocity range C}\\
1.364807& 4.7(0.5)& 11.09(0.01)& 13.00(0.00)& 13.73(0.08)& 13.18(0.01)&12.40(0.03) & 12.83(0.04)& 11.98(0.05)&    $\le12.00$&$\le$18.62\\
1.364897& 2.5(0.7)& 10.97(0.01)& 14.31(0.03)& 14.31(0.27)& 13.60(0.02)&12.62(0.10) & 12.72(0.07)& 11.71(0.25)&    $\le12.00$&$\le$18.34\\
1.364963& 4.2(0.6)& 11.75(0.01)& 13.41(0.09)& 14.46(0.47)& 14.14(0.02)&13.40(0.04) & 12.73(0.07)& 12.63(0.01)& 12.10(0.07)  &$\le$18.57\\
1.365025& 2.9(0.4)& 11.21(0.01)& 14.16(0.18)& 15.06(0.49)& 13.87(0.02)&13.12(0.06) & 12.71(0.07)& 12.55(0.03)& 11.97(0.11)  &$\le$18.41\\
\multicolumn{2}{c}{Total}&11.97(0.01)&14.59(0.07)&15.23(0.34)&14.43(0.01)&13.65(0.03)&13.35(0.03)& 12.97(0.02)& 12.35(0.08)&$\le$19.26\\
\\
\multicolumn{11}{c}{Velocity range D}\\
1.365116& 5.4(0.7)& 11.14(0.01)& 13.42(0.06)& 13.87(0.08)& 13.62(0.01)&12.93(0.06) & 12.26(0.16)& 12.43(0.02)& 12.34(0.02)  &$\le$18.69 \\
1.365217& 6.8(0.7)& 11.38(0.01)& 14.16(0.80)& 14.14(0.05)& 14.06(0.01)&13.25(0.04) & 12.63(0.08)& 12.57(0.01)&    $\le12.00$&$\le$18.78 \\
1.365305& 3.9(0.4)& 10.78(0.02)& 13.99(0.02)& 13.68(0.08)& 13.83(0.02)&12.69(0.07) & 12.04(0.24)& 11.96(0.04)& 11.96(0.05)  &$\le$18.54 \\
\multicolumn{2}{c}{Total}&11.64(0.01)&14.43(0.43)&14.41(0.04)&14.35(0.01)&13.50(0.03)&12.85(0.07)&12.86(0.01)&12.49(0.03)   &$\le$19.31 \\
\\
\multicolumn{11}{c}{Velocity range E}\\

1.365430& 5.4(0.3)& 10.70(0.02)& 13.20(0.00)& 13.42(0.08)& 13.37(0.01)&12.51(0.01) & 12.28(0.15)  & 11.95(0.06)   &  $\le12.00$ &$\le$18.69\\
1.365570& 3.9(0.2)& 10.48(0.03)& 13.69(0.01)& 13.37(0.09)& 13.52(0.01)&12.52(0.01) & 12.11(0.19)  & $\le11.70$    &  $\le12.00$ &$\le$18.55\\
1.365667& 2.5(0.6)& 10.43(0.04)& 12.74(0.00)& 13.37(0.10)& 13.04(0.01)&12.19(0.03) &  $\le12.00$  & $\le11.70$    &  $\le12.00$ &$\le$18.35\\
1.365759& 3.3(0.4)& 10.43(0.04)& 13.11(0.01)& 13.19(0.10)& 13.10(0.01)&12.24(0.02) & 11.64(0.54)  & $\le11.70$    &  $\le12.00$ &$\le$18.46\\
\multicolumn{2}{c}{Total}&11.12(0.01)&13.92(0.01)&13.95(0.05)&13.90(0.01)&12.99(0.01)& 12.56(0.14)&....           &....         &$\le$19.29\\
\\
\multicolumn{11}{c}{Velocity range F}\\
1.365906& 8.0(0.2)&  $\le9.47$    & 12.21(0.01)& 13.38(0.11)   & 12.65(0.02)   &$\le11.50$	 & $\le12.00$    & $\le11.70$    &  $\le12.00$ &$\le$18.86\\
1.366008& 1.8(0.1)& 10.08(0.09)   & 12.13(0.01)& 12.72(0.31)   & 12.30(0.05)   &$\le11.50$	 & $\le12.00$    & $\le11.70$    &  $\le12.00$ &$\le$18.21\\
1.366197& 3.5(0.4)&  $\le9.47$    & 11.48(0.02)&$\le13.0$      &$\le12.00$     &$\le11.50$	 & $\le12.00$    & $\le11.70$    &  $\le12.00$ &$\le$18.49\\
1.366330& 1.9(0.7)&  $\le9.47$    & 11.23(0.02)&$\le13.0$      &$\le12.00$     &$\le11.50$       & $\le12.00$    & $\le11.70$    &  $\le12.00$ &$\le$18.23\\
\hline
\multicolumn{11}{l}{Velocity ranges are defined in Fig.~1.}\\
\multicolumn{11}{l}{$^1$ Column densities are given in logarithmic units.}\\
\multicolumn{11}{l}{$^2$ upper limit on $N$(H~{\sc i}) obtained from the GMRT 
spectrum assuming $T_{\rm S}=100$ K; for higher $T_{\rm S}$ 
this should be multiplied by ($T_{\rm S}$(K)/100)}\\
\multicolumn{11}{l}{$b$ blended with other lines}\\ 
\end{tabular}
}
\label{table1}
\end{center}
\end{table*}

\begin{figure}
\begin{center}
\psfig{figure=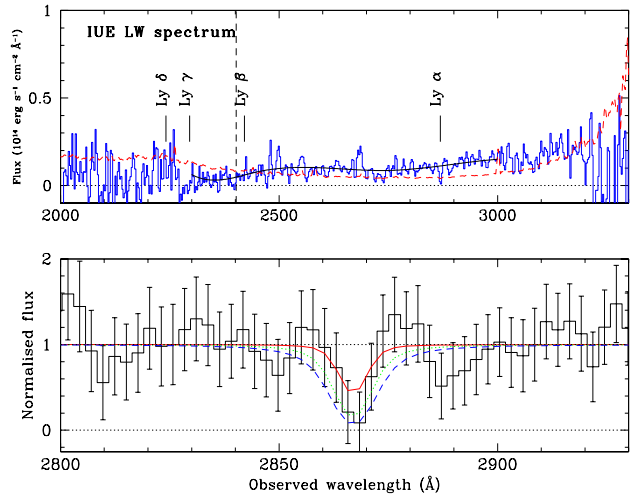,height=7cm,width=9.0cm,angle=270}
\caption[]{Top panel shows the low resolution ($\simeq 5.5$~\AA) IUE spectrum of 
PKS 0237$-$233, the fitted continuum (smooth continuous curve) and the error spectrum 
(dashed line). The expected positions of the Lyman series lines at \zabs = 1.3647 are 
marked by vertical lines. Single component Voigt profile fits
for log $N$~(H~{\sc i}) = 19, 19.5 and 20.0 are shown by, respectively, continuous, 
dotted and dashed profiles. These fits suggest that the system is a sub-DLA with
log $N$(H~{\sc i}) $\le$ 20.0.
}
\label{fig3}
\end{center}
\end{figure}

\begin{figure}
\begin{center}
\psfig{figure=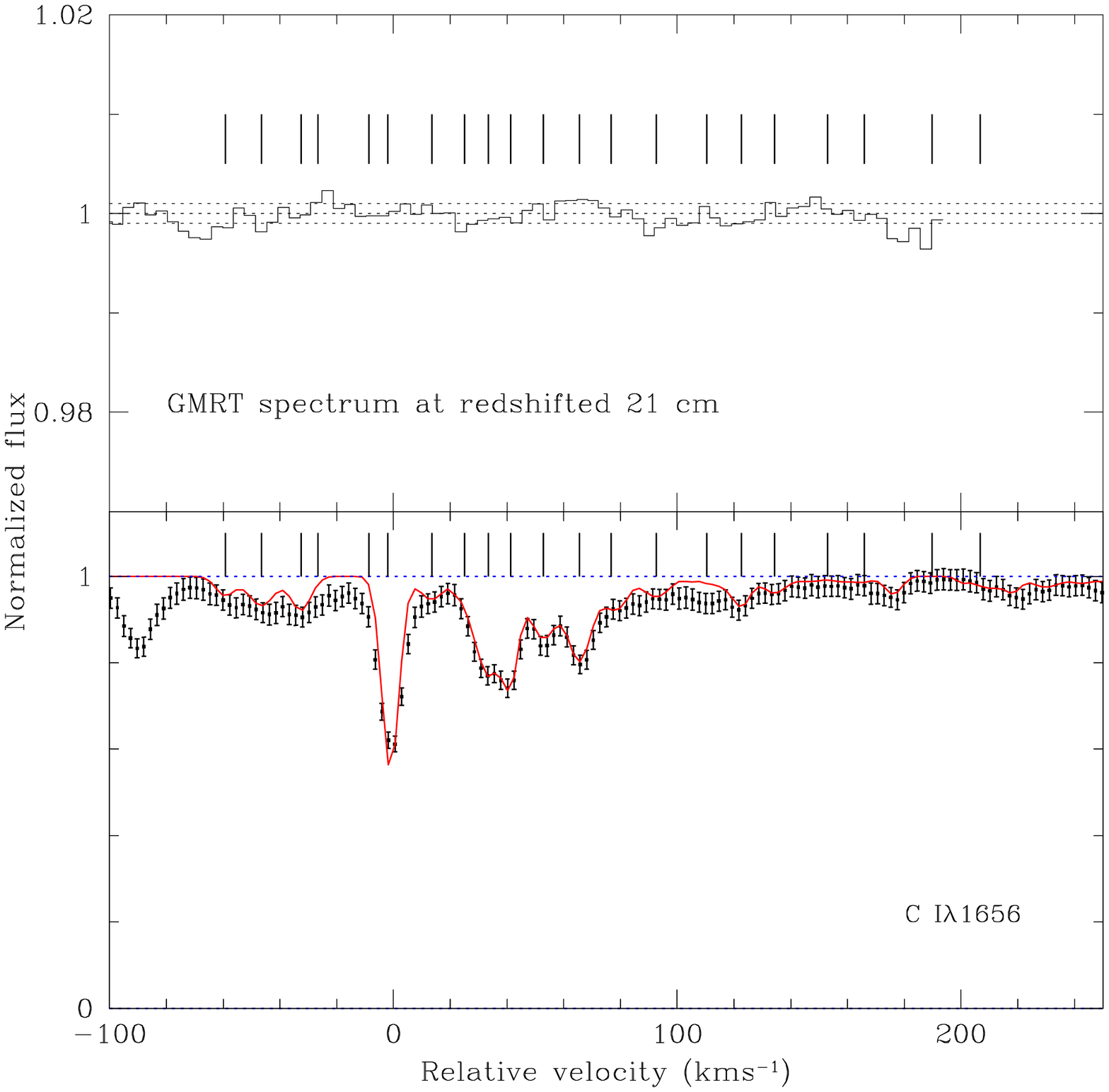,height=7cm,width=9.0cm,angle=0}
\caption[]{
Top panel shows the GMRT spectrum on a velocity scale
centered at \zabs~=~1.3647. The horizontal dashed lines correspond to
the mean and 1$\sigma$ range in the normalised flux.
The corresponding C~{\sc i}$\lambda$1606 absorption is plotted
in the bottom panel. Vertical tick marks show
the location of different velocity components derived from
the metal line profiles.
}
\label{fig4}
\end{center}
\end{figure}

%We detect a number of metal lines at \zabs = 1.3647 in the UVES spectrum. 
The velocity profiles of some of the neutral and singly ionized species seen
at \zabs = 1.3647 together with a multicomponent Voigt profile fit are shown in Fig~\ref{fig1}.
They are spread over $\simeq 300$ \kms.
%The profile of the Mg~{\sc ii} absorption is spread over $\simeq 300$ \kms.
The absorption lines of  Mg~{\sc i}, Si~{\sc ii}, Al~{\sc ii}, Al~{\sc iii} and
Fe~{\sc ii} fall in the spectral range with the best signal-to-noise ratio.
We also detect O~{\sc i}$\lambda$1302 in the blue part of the UVES
spectrum that has a lower SNR. Weak Mn~{\sc ii} lines are clearly 
detected in the strongest Mg~{\sc ii} components corresponding to 
velocity ranges B and C (see left bottom panel in Fig.~1). We also tentatively 
detect Cr~{\sc ii} in the same components (see Fig.~\ref{fig1}).

We detect C~{\sc i}, C~{\sc i$^*$} and C~{\sc ii$^*$} absorption in
most of the components (see Fig.~\ref{fig2}). Simultaneous Voigt profile
fitting of all the metal transitions requires 21 independent components.
The results are summarised in Table.~\ref{table1}.
To facilitate the discussion, we divide the velocity profiles into
5 distinct velocity regions marked as A, B, C, D and E in the bottom left
panel of Fig.~\ref{fig1}. The integrated column densities of different species 
within these velocity ranges are reported in Table~{\ref{table1}}.
Ni~{\sc ii}$\lambda$1709, Co~{\sc ii}$\lambda$1466,
Na~{\sc i}$\lambda\lambda$3303.3,3303.9, Ca~{\sc ii}$\lambda\lambda$3934,3969
are not detected, nor are the fine-structure lines of O~{\sc i} and Si~{\sc ii}. 

\subsection{Constraints on the total $N$(H~{\sc i})}

We used the IUE LW spectrum from the IUE archive to derive the 
total H~{\sc i} column density, $N$(H~{\sc i}). We detect a line 
at the expected position of the redshifted \lya absorption with rest equivalent width 
$3.0\pm1.0$\AA~ (Fig.~\ref{fig3}). If we assume the absorption to be 
in the damped portion of the curve of growth we derive log~$N$(H~{\sc i}) 
= 19.30$\pm$0.30. This means that $N$(H~{\sc i}) in individual components 
cannot be higher than this value. If the \lya line is the blend of saturated 
\lya absorptions from individual metal-line components, then the total $N$(H~{\sc i}) could 
be less than this value. 

The derived constraint on $N$(H~{\sc i}) is consistent with ASCA \& ROSAT 
observations that do not detect excess soft X-ray absorption on top of 
the galactic absorption (Yamasaki et al. 1998). Thus, the allowed range 
in $N$(H~{\sc i}) is very much consistent with the system being either a 
sub-DLA or a high column density Lyman Limit system. 
%Rough estimate 
Upper limits on $N$(H~{\sc i}) in the individual components can be obtained 
from our GMRT spectrum for an assumed spin-temperature ($T_{\rm S}$). 
The GMRT spectrum is shown in the upper panel of Fig.~\ref{fig4}. Despite 
reaching a 1$\sigma$ limit of 10$^{-3}$ for $\tau$(21 cm) we do not 
detect any absorption at the expected positions of the different individual 
components detected in the UV spectrum of neutral and/or singly ionized species. 

The neutral hydrogen column density in a velocity interval $v$ and $v+$d$v$ for an 
optically thin cloud is related to the optical depth, $\tau({\rm v})$, and to the
spin temperature ($T_{\rm S}$) by
\begin{equation}
N{\rm(H~I)}=1.835\times10^{18}~T_{\rm S}\int~\tau(v)~{\rm d}v~{\rm cm^{-2}}
\label{eq1}
\end{equation}
If we assume the absorption line to have a perfect Gaussian profile with
a peak optical depth $\tau_{\rm P}$, then the above equation will
become, 
\begin{equation}
N{\rm (H~I)}=1.93\times10^{18}~\tau_{\rm P}~T_{\rm S}~\Delta V~ {\rm cm^{-2}}
\label{eq2}
\end{equation}
where ${\Delta V}=2\sqrt{\rm ln~2}\times b$ is the FWHM of the Gaussian profile with 
$b$ the Doppler parameter. 

Note that the above equations assume the absorbing cloud to be homogeneous
and to cover the background source completely.  PKS 0237$-233$ is a very 
compact radio source that is not resolved in our GMRT 
observations. VLBA observations of this source at 2.32 and 8.55 GHz have
revealed structures at milli-arcsec scales (Fey et al. 1996). Most of 
the flux is in two components that are roughly separated by 
$\sim$ 10 milli-arcsec (see also  Fomalont et al. 2000). These components 
account for all the flux at 2.32 and 8.55 GHz implying there is no broad
diffuse radio emission in this source.
The strongest of these components has 
~60\% of the flux at 2.32 GHz (Fey et al 1996) and 70\% of the 
flux at 5 GHz (Famalont et al 2000). 
It has therefore a steep spectrum
($f_\nu\propto\nu^{-1.2}$) and probably dominates the flux at 
610 MHz. The second brightest component has a flat spectrum 
($f_\nu\propto\nu^{-0.2}$) and should correspond to
the core of the radio source and therefore to the optical counterpart. 
If this is the case, then the optical and 610~MHz radio lines of sight
are separated by 10 milli-arcsec.
However, the spectral indices derived above are highly uncertain due to different 
beam sizes and component decompositions of the two above observations. 
Because of this, it is unclear which of these components is indeed associated 
with the optical point source and the angular separation of 10 milli-arcsec
should be considered as an upper limit.
One milli-arcsec at $z$ = 1.35 corresponds to a 
linear size of 8.5 pc (for a flat universe with $\Omega_\Lambda = 0.73$, 
$\Omega_m = 0.27$, H$_0$ = 71 km/s/Mpc). This means the optical
and radio sightlines are either coincident or separated by $\sim$85 pc. 

Assuming that the velocity width of the H~{\sc i} feature is the  
same as that of metals, 
and using $\tau_{\rm P}(3\sigma)=0.003$  and $T$ = 100 K, as expected in the cold 
neutral medium, we derive upper limits on $N$(H~{\sc i}) in individual 
components. These values are listed in last column of Table.~\ref{table1}. 
These limits should be scaled up by $T_{\rm S}(K)$/100.
Also for temperatures typical of the  warm neutral medium (WNM) or 
warm ionized medium (WIM) of our Galaxy,
higher $b$ values (say 9 \kms) should be considered.

It is clear that the uncertainties on the temperature and the
covering factor of the cloud in front of the radio-source prevents
any rapid conclusion. To go further in the analysis of this system,
in the following we construct photoionization  models and constrain them
using information from the 21~cm, \lya , C~{\sc i}, C~{\sc i}$^*$ 
and C~{\sc ii$^*$} absorption lines.

\begin{table}
\caption{Total column density:}
\begin{center}
\begin{tabular}{lc}%cc}
\hline
Species & log $N$(cm$^{-2}$)\\
\hline
\hline
 N(H~{\sc i}) &19.30$\pm$0.30\\
 N(O~{\sc i}) &$\ge$15.46    \\
 N(Mg~{\sc i})&12.39$\pm$0.00\\
 N(Mg II)     &$\ge$14.98    \\
 N(C II*)     &13.73$\pm$0.02\\
 N(Fe II)     &14.13$\pm$0.01\\
 N(Si II)     &14.89$\pm$0.01\\
 N(Al II)     &$\ge 13.46$   \\
 N(Al III)    &13.53$\pm$0.01\\
 N(Mn II)     &11.81$\pm$0.03\\
 N(Cr II)     &12.32$\pm$0.13\\
\\
\hline
\end{tabular}
\end{center}
\label{avemet}
\end{table}

\section{Model calculations}
\label{model}
\begin{figure}
\psfig{figure=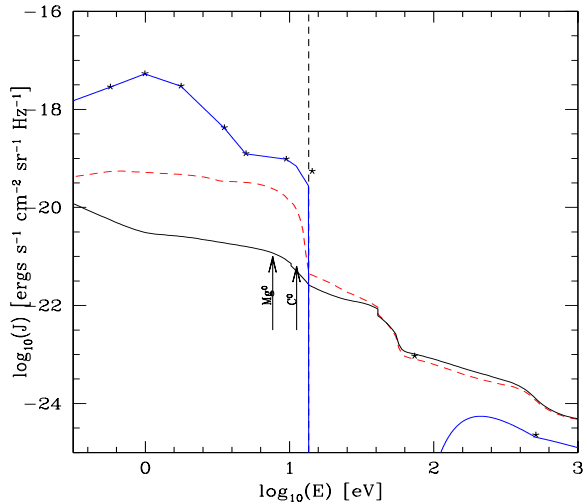,height=7.cm,width=8.0cm,angle=0}
\caption[]{Input spectral energy distributions (SEDs) used in our model calculations.
Continuous and dashed lines are SED taken from Haardt \& Madau (2005)
for the UV extra-galactic background contributed respectively by QSOs alone and QSOs and
Lyman break galaxies. Stars indicate the Galactic UV background SED
computed by Black (1987). The vertical dashed line shows the Lyman Limit;
the ionization potentials of C$^0$ and Mg$^0$ are shown with arrows.}
\label{sed}
\end{figure}
\begin{figure*}
\begin{center}
\psfig{figure=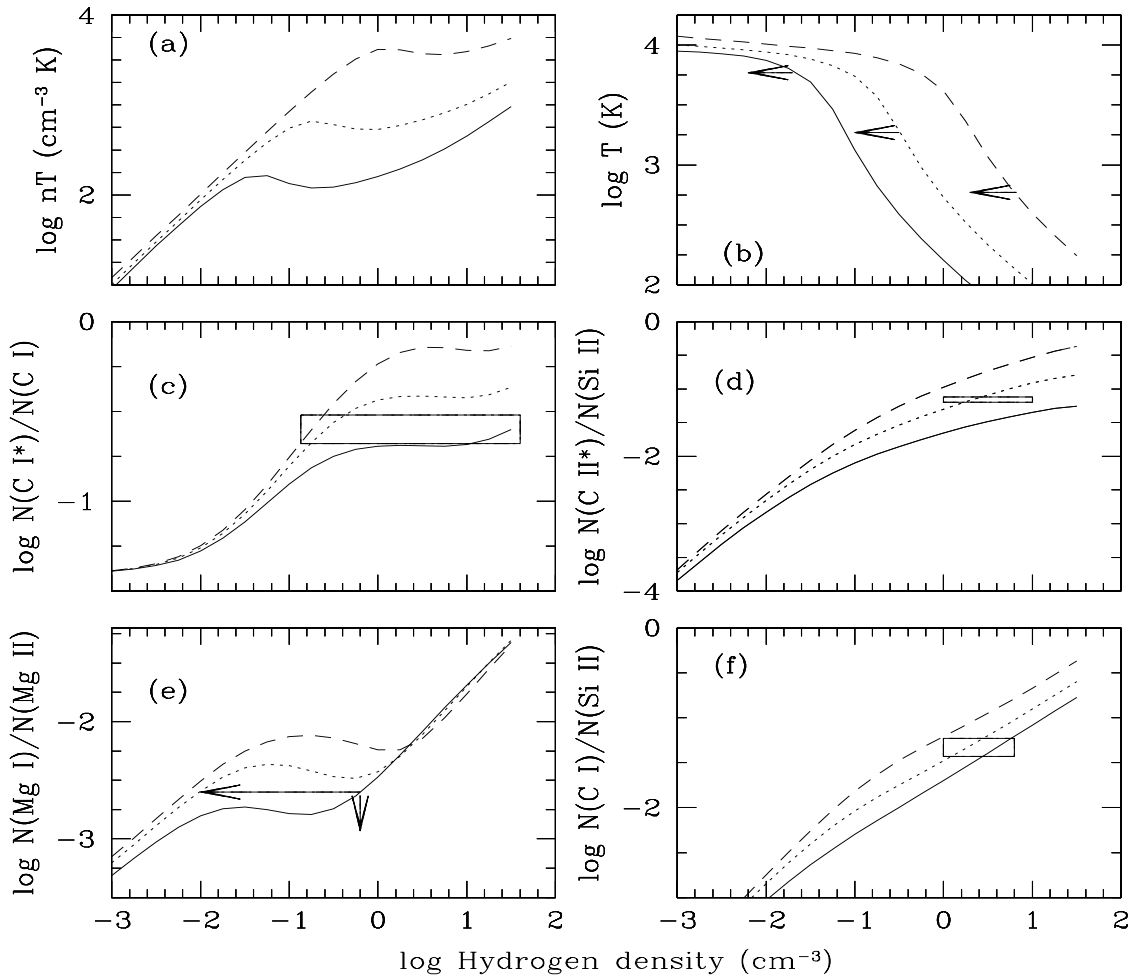,height=14cm,width=16cm,angle=0}
\caption[]{Results of photoionization models. The solid, short-dashed and
long-dashed curves are for, respectively, 
log $N$(H~{\sc i}) = 20.0, 19.5 and 19.0. Metallicity is 0.5 Z$_\odot$.
Constraints from the observed ratios of column densities integrated along the line of 
sight are shown in each panel as boxes or arrows. The arrows
in panel (b) indicate the upper limits on $n_{\rm H}$ 
obtained from the upper limit on the 21-cm optical depth
assuming the spin temperature to be equal to the kinetic temperature and
complete coverage of the background radio source.
}
\label{fig5}
\end{center}
\end{figure*}

\begin{figure*}
\begin{center}
\psfig{figure=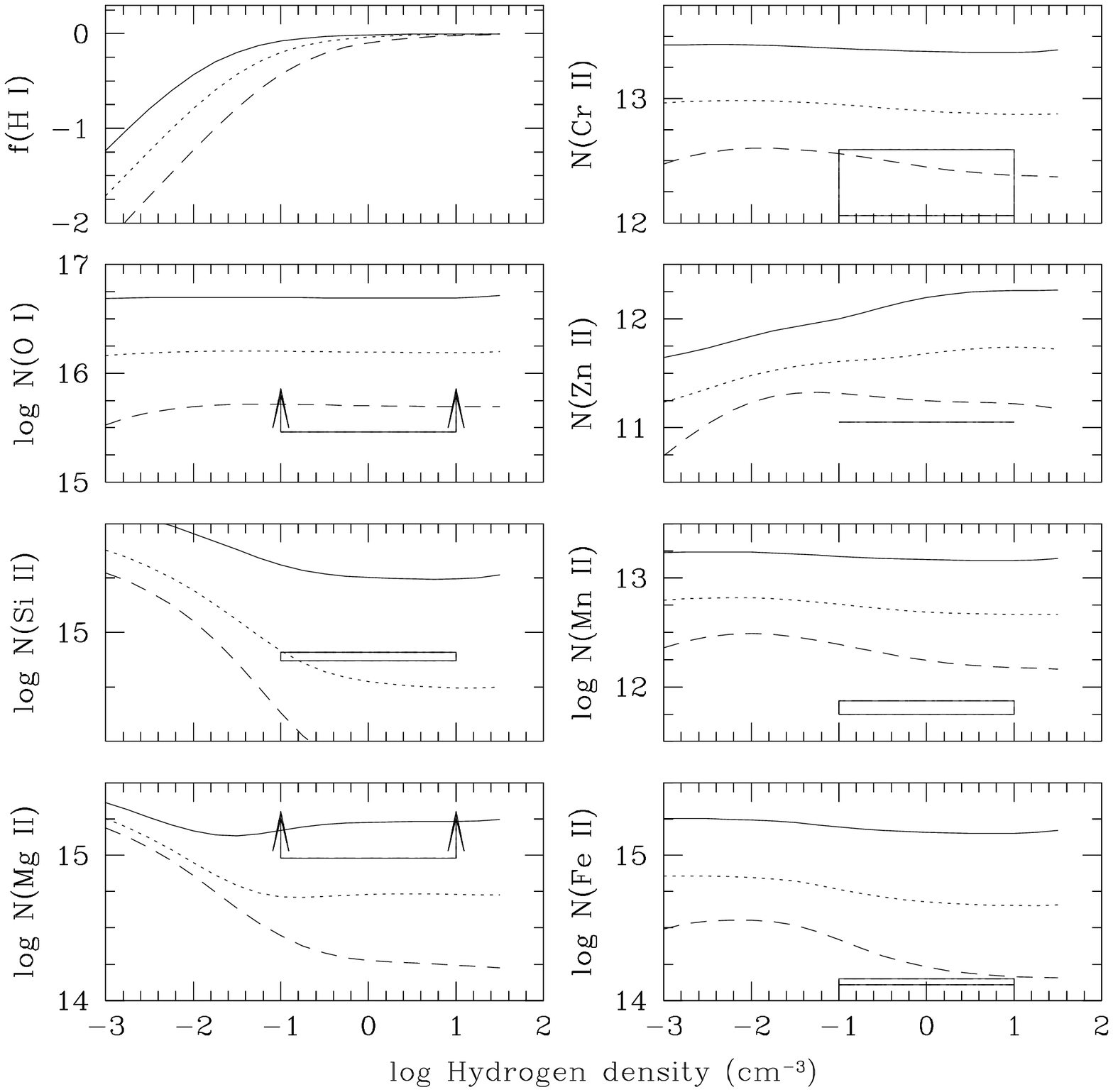,height=15cm,width=16.0cm,angle=0}
\caption[]{
Results of photoionization models. Various ion column densities
predicted by the models are plotted as a function of hydrogen density.
Symbols are as in Fig.~\ref{fig5}. $f$(H~{\sc i})~=~$N$(H~{\sc i})/$N$(H).
}
\label{fig6}
\end{center}
\end{figure*}

As the total $N$(H~{\sc i}) in this system is less than 10$^{20}$ 
cm$^{-2}$, it is most likely that ionization correction is important 
when deriving metallicity for most of the species. As an exception to 
this, O~{\sc i} and H~{\sc i} are coupled through charge-exchange reaction 
and for $N$(H~{\sc i})$\ge 10^{19}$ cm$^{-2}$, $N$(O~{\sc i})/$N$(H~{\sc i}) 
directly reflects [O/H] (see Viegas, 1995 and results presented below). 
As O~{\sc i}$\lambda$1302 is possibly saturated in the central components
(see Fig.~\ref{fig1}) we can derive only a lower limit on $N$(O~{\sc i}).
We find [O/H]$\ge-0.33$ if we consider log $N$(H~{\sc i})$\simeq$ 19.3. 
High metallicity is also inferred from Mg~{\sc ii} and Si~{\sc ii} 
(see Table.~\ref{avemet}). Clearly the gas under consideration must have 
gone through (or is going through) a prolonged period of vigorous 
star-formation activity. 

To correct for ionization we use model calculations 
performed with the photoionization-simulation code CLOUDY (Ferland et al. 1998). 
Keeping in mind the preferred value log~$N$(H~{\sc i})~=~19.3$\pm$0.30, we
consider three total H~{\sc i} column densities, 
log $N$(H~{\sc i}) = 19.0, 19.5 and 20.0, at which the calculation is 
stopped.
Relative metal abundances are considered solar and the total metallicity
is  Z = 0.5 Z$_\odot$. This is motivated by the above
inferred oxygen metallicity.

Most of the model calculations discussed below use a ionizing radiation 
spectrum that is a combination of  the meta-galactic UV background at 
$z = 1.35$ contributed by QSOs and galaxies as computed by Haardt \& Madau 
(2005) (dashed SED in Fig.~\ref{sed}) and the Cosmic Microwave Background with 
$T_{\rm CMBR}$ = 6.4 K. Relative to quasars, galaxies contribute approximately 
an order of magnitude higher flux in the energy ranges corresponding to the 
ionization potentials of C$^0$ and Mg$^0$ (See also Wolfe et al. 2004). 
We also consider models with additional local radiation (continuous curve 
with points in Fig.~\ref{sed}).  For this purpose we use the Black (1987) 
galactic background radiation after removing the H~{\sc i} ionizing photons 
as usually done in the case of ISM simulations (see e.g. Draine \& Bertoldi, 
1996; Shaw et al. 2006). We consider dust to metal ratio 1/10 of that 
observed in the Galactic ISM.   

First we mainly focus our attention on models with meta-galactic
radiation field contributed by QSOs and galaxies.
For a given values of total $N$(H~{\sc i}) and metallicity (Z)
we generate grids of models varying the density in the range, 
$-3\le{\rm log}~n_{\rm H}{\rm (cm^{-3})}\le1.5$. The results of our model
calculations are summarized in Figs.~\ref{fig5} and \ref{fig6}.
%Panels (a) and (b) in Fig.~\ref{fig5} respectively give the gas 
%pressure and temperature as a function of $n_{\rm H}$. 
%
%
%In panels (c) and (d) we plot, respectively, 
%log~$N$(C~{\sc i$^*$})/$N$(C~{\sc i}) and  
%log~$N$(C~{\sc ii$^*$})/$N$(Si~{\sc ii}) as a function of $n_{\rm H}$.
%This can be used to constrain the gas pressure.
%The ionization state of the gas can be constrained from 
%$N$(Mg~{\sc i})/$N$(Mg~{\sc ii}) and $N$(C~{\sc i})/$N$(Si~{\sc ii}) shown in 
%panels (e) and (f) of Fig.~\ref{fig5}. 
%
From these figures it can be seen that, for the range of 
$N$(H~{\sc i}) considered here, most of the constraints are 
consistent with 0$\le{\rm log[}n_{\rm H}~({\rm cm^{-3})]}\le$ 1. This 
density range corresponds to the gas being part of a CNM phase ($T<100$ K) only 
for the model with log $N$(H~{\sc i}) = 20 (see panel (a) of Fig.~\ref{fig5}).
In the other two models, that have $N$(H~{\sc i}) consistent with the observed range, 
the gas has the characteristics of a WNM phase or is in
the thermally unstable region. Therefore, it seems most likely that 
the C~{\sc i} absorption is originating either from WNM or WIM.
The $N$(Mg~{\sc i})/$N$(Mg~{\sc ii}) ratio predicted by these models
is larger than the observed ratio. This can be explained by
the large uncertainties in $N$(Mg~{\sc ii}) due to 
heavy saturation of the absorption lines (see Fig.~1).

The models allow us to estimate metallicity and depletion.
From Fig.~\ref{fig6} it is clear that for $-1\le{\rm log}~n_{\rm H} 
{\rm (cm^{-3})}\le1$ the ionization corrections for H~{\sc i}, O~{\sc i}, 
Mg~{\sc ii} and Si~{\sc ii} are negligible. Thus these ionization states 
can be used directly to infer the overall metallicity of the system. 
For $N$(H~{\sc i})$\simeq 10^{19.5}$ cm$^{-2}$ the observed column density 
of Si~{\sc ii} and the lower limits on O~{\sc i} and Mg~{\sc ii} column 
densities are consistent with Z = 0.56 Z$_\odot$. For such a metallicity 
the predicted column densities of Fe~{\sc ii} and Mn~{\sc ii} 
are about an order of magnitude higher than the observed column 
densities. The lower gas-phase abundance of Iron co-production elements 
with respect to the $\alpha-$process elements is confirmed by the 
measured low column density of Cr~{\sc ii} and the absence of Ni~{\sc ii} 
absorption. If this difference is due to depletion onto dust, then 
predictions for the Zn~{\sc ii} column density should match the observed 
value. This is not the case and Zn~{\sc ii} is overpredicted by the models. 
This argues toward an excess of $\alpha-$process elements.

Before drawing strong conclusions on the physical state of the gas, size of
the absorbing region and the metal enrichment scenario it may be interesting to
investigate the physical state of  C~{\sc i} sub-components. 
%Indeed, as the velocity spread of the C~{\sc i} absorption is large, our 
%single cloud approximation is probably not accurate. 
While single cloud models give the average property of the gas, modelling
individual velocity ranges will allow us to investigate variations of physical
conditions across the profile.
In the following 
Section we therefore discuss  properties of the different C~{\sc i} velocity ranges 
we have identified before (see Fig.~\ref{fig1}). In the corresponding models we have varied 
$N$(H~{\sc i}) from 10$^{18}$ to 10$^{19}$ cm$^{-2}$.

\subsection{Discussion on the different velocity ranges}

\begin{figure}
\psfig{figure=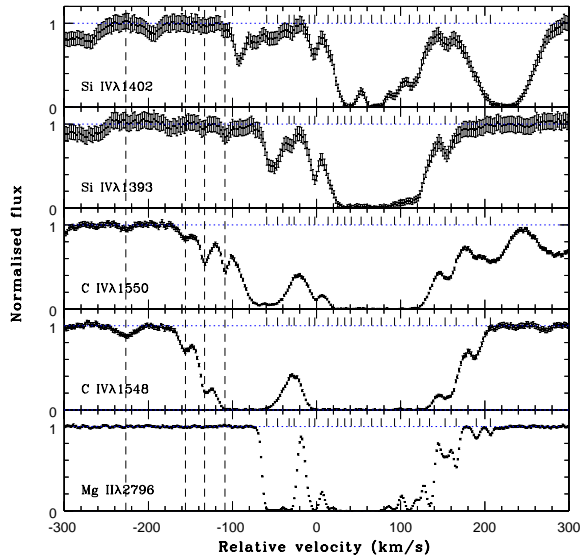,height=8cm,width=8.0cm,angle=0}
\caption{Profiles of high-ionization lines detected at \zabs = 1.3647.
Tick marks indicate the location of the different velocity components 
seen in the low-ionization profiles. Vertical dashed-lines mark the location
of components that are detected only in the high ionization species.}
\label{highion}
\end{figure}

\subsubsection{Velocity range A}

This velocity range has three well resolved components and an additional 
weak component in the red wing. C~{\sc i} absorption is possibly present 
in the three main components with little C~{\sc i$^*$} absorption. 
We find that the upper limit on $N$(C~{\sc i$^*$}) is consistent with 
${\rm log}~n_{\rm H} {\rm (cm^{-3})}\le0.0$ 
for models with log $N$(H~{\sc i})$\le$19. C~{\sc ii$^*$} absorption is 
blended with a strong Ly-$\alpha$ line at \zabs = 1.5. Therefore, there 
is no strong constraint on the density for this velocity range. This 
velocity range has log $N$(Mg~{\sc i})/$N$(Mg~{\sc ii}) $\simeq$ $-2.4$ and  
log $N$(O~{\sc i})/$N$(Si~{\sc ii}) = 0.06$\pm$0.05. In our model 
calculations these ratios are consistent with {log $N$(H~{\sc i}) ~$\le$18.5 and $-3\le$log$ n_{\rm H}$ (cm$^{-3}$)~$\le -2.0$. 
Such low densities are needed to explain that $N$(O~{\sc i})/$N$(Si~{\sc ii}) 
is different from [O/Si]. We also find that the observed log 
$N$(Al~{\sc iii})/$N$(Al~{\sc ii}) = $-0.22\pm0.11$ ratio is consistent 
with low $N$(H~{\sc i}) and low density.
The presence of possible additional local radiation below the Lyman
limit from the local stellar population would reproduce the observed
$N$(Mg~{\sc i})/$N$(Mg~{\sc ii}) ratio at higher densities 
(Fig.~\ref{fblack}). In any case models for this gas require large 
ionization corrections for H~{\sc i} (see Fig.~\ref{fig6}). We therefore 
do not attempt to further infer 
metallicities and depletion pattern in this velocity range. 

\subsubsection{Velocity range B}

\begin{table}
\caption{Model results for the velocity range B}
\begin{tabular}{lccc}
\hline
Species & log $N$ & \multicolumn{2}{c}{log $N$(H~{\sc i}) = 19.0$-$18.5}\\
        & (Observed)  & Z = 0.1 Z$_\odot$ & Z = 0.5 Z $_\odot$ \\
\hline
\hline
C~{\sc i}     &12.92$\pm$0.01& 11.99$-$12.14& 12.66$-$12.84\\
C~{\sc i$^*$} &12.56$\pm$0.02& 11.28$-$11.60& 11.95$-$12.30\\
C~{\sc ii$^*$}&13.30$\pm$0.03& 12.14$-$12.38& 12.81$-$13.06\\
O~{\sc i}$^a$ &14.23$\pm$0.05& 14.24$-$14.72& 14.94$-$15.42\\
Mg~{\sc i}    &11.37$\pm$0.01& 11.50$-$11.74& 12.08$-$12.31\\
Mg~{\sc ii}$^a$&13.50$\pm$0.01& 13.40$-$13.77& 14.06$-$14.44\\
Al~{\sc ii}   &12.22$\pm$0.01& 12.41$-$12.85& 13.05$-$13.47\\
Si~{\sc ii}   &13.77$\pm$0.01& 13.48$-$13.93& 14.30$-$14.45\\ 
Fe~{\sc ii}   &13.07$\pm$0.01& 13.30$-$13.70& 13.99$-$14.42\\
Mn~{\sc ii}   &10.93$\pm$0.07& 11.25$-$11.67& 11.98$-$12.39\\
Zn~{\sc ii}   &10.61$\pm$0.20& 10.34$-$10.66& 10.99$-$11.30\\
Cr~{\sc ii}   &$\le11.30$$^b$& 11.37$-$11.80& 12.16$-$12.56\\
\hline
\multicolumn{4}{l}{$^a$ probably saturated; $^b$ 3$\sigma$ upper limit.}\\
\end{tabular}
\label{tab2}
\end{table}

This velocity range has two components (one main component and
a weak blend in the blue wing) and is associated with the strongest C~{\sc i} 
and  C~{\sc ii$^*$} absorptions. From Fig.~\ref{fig1} it appears that the O~{\sc i} absorption is not
heavily saturated. This velocity range contributes about $\sim$20\% of
the total $N$(O~{\sc i}). This means that about 20\% of the total
$N$(H~{\sc i}) is in this velocity range and $N$(H~{\sc i}) is probably not  
much higher than  10$^{19}$ cm$^{-2}$. The observed 
log $N$(O~{\sc i})/$N$(Si~{\sc ii}) = 1.03$\pm$0.13 and  
log $N$(O~{\sc i})/$N$(Mg~{\sc ii}) = 1.30$\pm$0.16 ratios are consistent
with solar metallicity ratios [O/Si]$_\odot$ = 1.13 and [O/Mg]$_\odot$ = 1.11. 
As all these species are $\alpha-$process elements it is most likely 
that ionization corrections are negligible.

The measured $N$(C~{\sc i$^*$})/$N$(C~{\sc i}) is consistent with 
$-1.0\le{\rm log [}n_{\rm H}({\rm cm}^{-3})]\le-0.6$ for 
18~$\le$~log [$N$(H~{\sc i}) (cm$^{-2}$)]~$\le$~19.5.
The temperature of the gas is higher than 1000 K for the range in metallicity 
and log~$N$(H~{\sc i}) considered here.
%This is consistent with the upper limit on $T_{\rm S}$ derived from 
%the absence of 21-cm in our GMRT spectrum.
The gas pressure in this component is consistent with 
$3.0\le{\rm log[P/k (cm^{-3} K)]} \le 3.6$. The gas is therefore warm with a neutral 
hydrogen fraction $\ge$0.7. 

The observed $N$(C~{\sc ii$^*$})/$N$(Si~{\sc ii}) ratio is much higher than 
that predicted by the models in the range of $n_{\rm H}$ derived  
from $N$(C~{\sc i$^*$})/$N$(C~{\sc i}).
The difference is higher for higher $N$(H~{\sc i}) models. 
In Table.~\ref{tab2} 
we give the column densities of different species predicted in our model for
$-1.0\le{\rm log}[n_{\rm H}({\rm cm}^{-3})]\le-0.6$, 
$18.5 \le $~log~[$N$(H~{\sc i})~(cm$^{-3}$)]~$\le 19$ and 
two different assumed values of metallicities 
(Z = 0.1 and 0.5 Z$_\odot$). From this table it is clear that 
the observed column densities of Si~{\sc ii}, Mg~{\sc ii}, Zn~{\sc ii} 
and O~{\sc i} are well reproduced by the models with Z = 0.1 Z$_\odot$. 
However, the observed column densities of C~{\sc i}, C~{\sc i$^*$} and C~{\sc ii$^*$} 
are higher than in the models. Also the well measured 
column densities of Fe~{\sc ii} and Mn~{\sc ii} are a factor of two below the 
model predictions. Models with Z~$\ge$~0.5~Z$_\odot$ are needed to
explain $N$(C~{\sc i}) and $N$(C~{\sc ii$^*$}). These models will
require a depletion of O, Mg, Al and Si by up to 0.6 dex and  Fe, Mn and Cr
by up to 1.0 dex. 

The observed ratio, log $N$(Al~{\sc iii})/$N$(Al~{\sc ii})~=~$-0.15\pm0.17$, 
is similar to what is observed for component A. This clearly suggests
the presence of low-density ionized gas (with low column density) 
over the same velocity range as the C~{\sc i} gas. This is confirmed by
the presence of strong Si~{\sc iv} and C~{\sc iv} absorption lines
with similar velocity structure (see Fig.~\ref{highion}).
It is interesting to note that the observed $N$(Mg~{\sc i})/$N$(Mg~{\sc ii}) ratio
is explained by the models discussed here. This means
that there is no excess of UV radiation with energies less than 13.6 eV
in these velocity components (see Fig.~\ref{fblack}). In conclusion this 
velocity range seems to be well apart from any local star-forming 
region, has metal enrichment between 0.1 and 0.5~Z$_{\odot}$ and reasonable depletion of 
iron co-production elements.

\begin{figure*}
\begin{center}
\psfig{figure=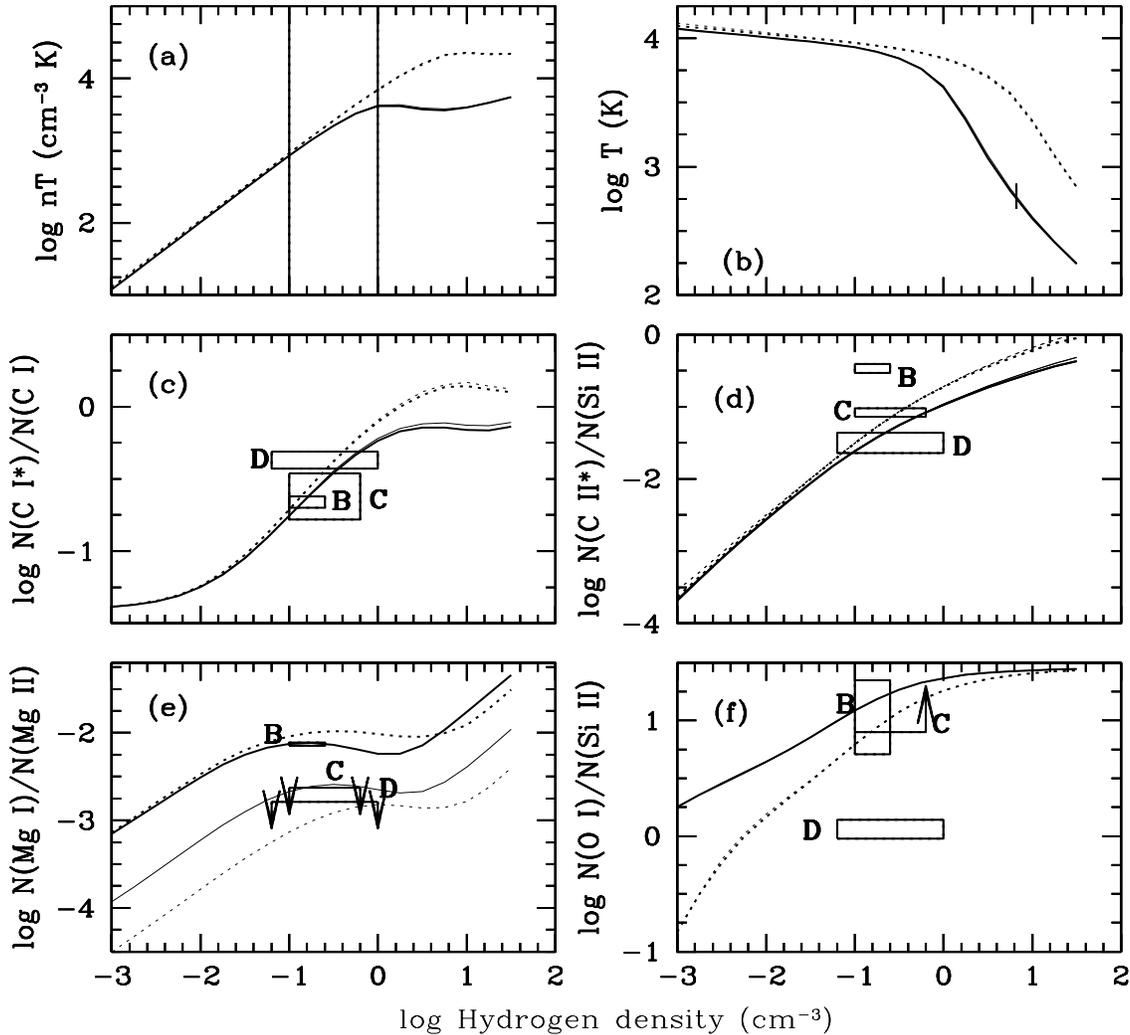,height=15cm,width=16.0cm,angle=0}
\caption[]{
Results of photoionization models with an excess of local radiation in addition
to the meta-galactic UV background radiation. Solid and dashed lines are for,
respectively, log~$N$(H~{\sc i}) = 19.0 and 18.5. Thick and thin curves
are for models with and without local radiation added to the
meta-galactic background radiation.
}
\label{fblack}
\end{center}
\end{figure*}

\subsubsection{Velocity range C}
\begin{table}
\caption{Model Results for the velocity range C}
\begin{tabular}{lccc}
\hline
Species & $N$ & \multicolumn{2}{c}{log $N$(H~{\sc i}) = 19.0$-$18.5}\\
        & (Observed)  & Z = 0.1 Z$_\odot$ & Z = 0.5 Z $_\odot$ \\
\hline
\hline
C~{\sc i}      &12.97$\pm$0.02 & 12.04$-$12.14& 12.66$-$12.88\\
C~{\sc i$^*$}  &12.65$\pm$0.10 & 11.28$-$11.98& 11.94$-$12.54\\
C~{\sc ii$^*$} &13.35$\pm$0.03 & 12.14$-$12.47& 12.81$-$13.13\\
O~{\sc i}$^a$  &   $\ge15.23$  & 14.22$-$14.72& 14.92$-$15.42\\
Mg~{\sc i}     &11.97$\pm$0.01 & 11.43$-$11.74& 11.99$-$12.31\\
Mg~{\sc ii}$^a$&   $\ge14.59$  & 13.30$-$13.77& 13.98$-$14.44\\
Al~{\sc ii}    &13.08$\pm$0.01 & 12.29$-$12.85& 12.95$-$13.47\\
Si~{\sc ii}    &14.43$\pm$0.01 & 13.37$-$13.93& 14.04$-$14.45\\
Ca~{\sc ii}    & $\le12.00^b$  &              & 12.45$-$12.68\\ 
Mn~{\sc ii}    &11.53$\pm$0.04 & 11.20$-$11.67& 11.91$-$12.39\\
Fe~{\sc ii}    &13.65$\pm$0.03 & 13.23$-$13.70& 13.90$-$14.42\\
Ni~{\sc ii}    & $\le12.95^b$  &              & 12.76$-$13.34\\
Zn~{\sc ii}    & $\le11.10^b$  & 10.28$-$10.66& 10.94$-$11.30\\
Cr~{\sc ii}    & 11.63$\pm$0.14& 11.34$-$11.80& 12.11$-$12.56\\
\hline
\multicolumn{4}{l}{$^a$ probably saturated; $^b$ 3$\sigma$ upper limit.}\\
\end{tabular}
\label{tab3}
\end{table}

This velocity range has $\ge 60\%$ of the total O~{\sc i} column density
distributed in 4 distinct velocity components. Only two of them show
detectable C~{\sc i$^*$} absorption and other two components are 
mainly in the wings of the absorption. Clearly most of the neutral
hydrogen in the system must originate from
this velocity range. The total column densities of C~{\sc i}, C~{\sc i$^*$} 
and C~{\sc ii$^*$} in this velocity range are consistent with that seen for 
velocity range B. This means that the average pressure in both the velocity 
ranges are more or less consistent with one another. The measured 
$N$(C~{\sc i$^*$})/$N$(C~{\sc i}) ratio is consistent with 
$-1.0\le{\rm log} [n_{\rm H}({\rm cm}^{-3})]\le-0.2$ for the range of 
$N$(H~{\sc i}) considered in our models.
We also find that the models with 
18.5~$\le$log [$N$(H~{\sc i})(cm$^{-2}$)] $\le$~19 will also
reproduce the observed $N$(C~{\sc ii$^*$})/$N$(Si~{\sc ii}) ratio.
This means that we do not require enhancement of C with
respect to Si and Mg as in component B.  The gas pressure in this velocity range
is $3.0\le{\rm log}[P/{\rm k (cm^{-3} K)]} \le 3.8$. Like in velocity
range B, the C~{\sc i} absorption from this component comes
from a WIM or WNM.

O~{\sc i} and Mg~{\sc ii} absorptions are saturated and the 
derived column densities of Si~{\sc ii}, Al~{\sc ii}, Fe~{\sc ii}, 
Mn~{\sc ii} and Cr~{\sc ii} are higher than in velocity range B. 
The column densities of C~{\sc i}, C~{\sc ii$^*$}, O~{\sc i}, Si~{\sc ii} 
and Mg~{\sc ii} are approximately consistent with  Z $\ge$ 0.5 Z$_\odot$ (see Table~\ref{tab3}). 
However the observed column densities of Fe~{\sc ii}, Mn~{\sc ii} and Cr~{\sc ii} and 
the 3 $\sigma$ upper limits on Ca~{\sc ii} and Ni~{\sc ii} are 
consistent with the iron co-production elements being under
abundant by $\simeq 0.4$ dex. 
This depletion of iron co-production elements with respect to $\alpha-$elements
is consistent with what is seen in the velocity range B. 

As we have only a lower limit on the Mg~{\sc ii} column density,
the actual $N$(Mg~{\sc i})/$N$(Mg~{\sc ii}) ratio may be smaller than 
that seen in velocity range B. From panel (e) in Fig.~\ref{fblack} it is 
clear that for log $N$(H~{\sc i}) = 19, the observed log $N$(Mg~{\sc i})/$N$(Mg~{\sc ii})
ratio is consistent with a UV radiation field as high as the galactic
mean UV field below 13.6 eV. However, the observed $N$(Mg~{\sc i}) and 
$N$(C~{\sc i}) are inconsistent with much larger radiation fields 
(of the order of 10 times the Galactic field) as required for the 
H$_2$ components in high-z DLAs.

Like in the case of velocity range B we notice that the Al~{\sc iii} column
density is similar to that of Al~{\sc ii} 
(log $N$(Al~{\sc iii})/$N$(Al~{\sc ii}) = $-0.04\pm0.02$). Also very strong 
Si~{\sc iv} and C~{\sc iv} absorptions are present with 
similar velocity structure as the low ionization lines (see 
Fig.~\ref{highion}). This again suggests that highly ionized gas is 
co-spatial with the C~{\sc i} component in this velocity range as well. 

\subsubsection{Velocity range D}
\begin{table}
\caption {Model results for range D}
\begin{tabular}{lcc}
\hline
Species & $N$ & \multicolumn{1}{c}{log $N$(H~{\sc i}) = 18.0}\\
        & (Observed)  & Z = 0.5 Z $_\odot$ \\
\hline
\hline
C~{\sc i}      &12.86$\pm$0.01 & 12.47$-$12.53\\
C~{\sc i$^*$}  &12.34$\pm$0.02 & 11.65$-$12.52\\
C~{\sc ii$^*$} &12.85$\pm$0.07 & 12.60$-$12.92\\
O~{\sc i}$^a$  &14.41$\pm$0.04 & 14.43\\
Mg~{\sc i}     &11.64$\pm$0.01 & 11.75$-$12.17\\
Mg~{\sc ii}$^a$&   $\ge14.43$  & 13.70$-$14.25\\
Al~{\sc ii}    &13.08$\pm$0.01 & 12.67$-$13.38\\
Si~{\sc ii}    &14.35$\pm$0.01 & 13.76$-$14.45\\
Fe~{\sc ii}    &13.50$\pm$0.03 & 13.59$-$13.97\\
\hline
\multicolumn{3}{l}{$^a$ probably saturated; $^b$ 3$\sigma$ upper limit.}\\
\end{tabular}
\label{tab4}
\end{table}

The O~{\sc i} absorption in this velocity range is not saturated. However,
Mg~{\sc ii} lines are saturated (see Fig.~\ref{fig1}). 
Only one out of the three components show detectable C~{\sc i$^*$} 
absorption. All components
have $N$(O~{\sc i})$\simeq$ $N$(Si~{\sc ii}) within 0.2 dex.
For models considered here, $N$(C~{\sc i$^*$})/$N$(C~{\sc i}) 
and $N$(C~{\sc ii$^*$})/$N$(Si~{\sc ii}) are consistent with,
$-1.2\le{\rm log [}n_{\rm H}({\rm cm}^{-3})]\le0.0$ and  
$2.7\le{\rm log[}P/{\rm k (cm^{-3} K)]} \le 3.8$.

Models predict log $N$(O~{\sc i})/$N$(Si~{\sc ii})$\ge$ 0.9 for
log~$N$(H~{\sc i}) $\ge$~18.5. Therefore observations are either consistent
with  log~$N$(H~{\sc i})$<$18.5 or require under-abundance of O with respect
to Si. It can be seen from Table~\ref{tab4}
that most of the column densities of singly ionized species can be
reproduced by models with log $N$(H~{\sc i}) = 18 and Z = 0.5 Z$_\odot$. 

\subsection{Discussion}

Rao \& Turnshek (2000) have shown that DLAs can be pre-selected based 
on the equivalent widths of Mg~{\sc ii}, Fe~{\sc ii} and Mg~{\sc i} absorption lines.
Specifically, they have found that 50\% of the absorbers with rest equivalent
width of Fe~{\sc ii}$\lambda$2600 and Mg~{\sc ii}$\lambda$2796 greater
than 0.5 \AA~ were confirmed DLAs. If they impose the condition 
$W$(Mg~{\sc i})$\ge$ 0.5 \AA~ then all the systems are confirmed DLAs.
It is also clear from Fig.~3.4 of Lane's PhD thesis (2000) that 50\% 
of these systems will be detected by 21-cm absorption.
The inferred low $N$(H~{\sc i}) from IUE data (log~$N$(H~{\sc i})~$\sim$~19.3)
and the absence of 21-cm absorption 
at \zabs = 1.3647 towards PKS 0237-233 are both consistent with the above findings 
as $W_{\rm r}$(Mg~{\sc i})$\simeq$0.3\AA~ in this system which is clearly a
sub-DLA. 
It is therefore somewhat surprising to have found a wide-spread C~{\sc i} absorption
associated with the system suggesting high metallicity. 
Wide-spread (i.e $\Delta V\ge100$ \kms) C~{\sc i}
absorption is seen at \zabs = 1.973 towards Q~0013$-$004 (Petitjean, Srianand \& 
Ledoux 2002) and \zabs = 1.962 towards Q~0551$-$366 (Ledoux, Srianand \& Petitjean 2002).
Molecular hydrogen is detected in these systems although
several of the C~{\sc i} components do not have 
associated H$_2$ absorption. C~{\sc i} absorption associated with
H$_2$ have been shown to originate mainly from CNM gas (Srianand et al. 
2005). However the nature of the gas in the components that do not 
have associated H$_2$ absorption is still unclear. 

In the case of the \zabs = 1.365 system toward PKS~0237$-$233 
a simple one cloud model can explain the total column densities.
Under this assumption of single cloud, it can be shown that
the gas producing C~{\sc i} absorption is most unlikely to be
a CNM gas if the actual $N$(H~{\sc i}) is close to that we measure from the IUE spectrum. 
Detailed modelling of individual velocity ranges are also consistent  with 
the gas responsible  for the C~{\sc i} absorption in this system being either
 WNM or WIM. In addition we also infer ionization and chemical differences
between individual velocity ranges.

Our GMRT observations give a good constraint on $N$(H~{\sc i})/$T_{\rm s}$ 
if we assume complete coverage of the background radio-source by the absorbing
gas. However, as pointed out before, the 
background radio source shows structures at milli-arcsec scale with a 
separation between the two brightest components of 85 pc at the absorption 
redshift. It is unclear which of these components coincides with the 
optical point source. If the brightest component that has most of the 
flux in the redshifted 21-cm range coincides with the optical source 
then we would have detected 21-cm absorption if the C~{\sc i} gas 
originates from the CNM. The non-detection would
be a definite argument for the gas being warm. 
However, if the weaker radio component coincides with the optical source, 
then even if the C~{\sc i} absorption originates from CNM gas our GMRT observations 
would not be able to detect the corresponding 21-cm absorption 
as the expected size of the cloud is smaller than the radio source separation.
It is known that strong Mg~{\sc ii} systems (with $W_{\rm r} > $0.3 \AA) show 
coherent absorption over more than $\sim 2$ kpc (Petitjean et al. 2000,
Ellison et al. 2004). The fact that we find 21 distinct components
along the line of sight to the optical point source means that it is
likely that the overall system covers both the radio components.
In that case our GMRT observations are consistent with 
no CNM with log~$N$(H~{\sc i})$\ge$ 19 along the radio source as well.
{Thus, it seems most likely that the region probed by the Mg~{\sc ii} 
absorption system at \zabs = 1.365 is composed mostly of a warm neutral 
medium and/or warm ionized medium}.

The inferred metallicities suggest that the 
gas has gone through an active star-formation period.
We do find the 
signature of strong $\alpha-$process element enhancement with
respect to Fe co-production elements. However, as we have not
detected the Zn~{\sc ii} absorption it is not clear, although
unlikely, whether the
difference could be attributed to dust-depletion. Photo-ionization models 
for the individual components are 
consistent with ionization by meta-galactic UV background.
However, inconsistent with very large local radiation field
as one sees in \h2 components in high-z DLAs. This suggests
that the system is, at the time of observation, in a low star-formation state.

Out of the 50 Mg~{\sc ii} systems studied in the ESO large programme
'The Cosmic Evolution of the IGM', only two systems show C~{\sc i} absorption 
(i.e \zabs = 1.365 and 1.6724 towards PKS 0237$-$233).  The only other 
sub-DLA towards radio quiet QSO HE 0001-2340 does not show
C~{\sc i} and other fine-structure lines (See Richter et al. 2005).
The inferred $N$(H~{\sc i}), metallicity,  and depletion pattern 
at \zabs = 1.365 toward PKS~0237$-$233 are similar to what is observed
at \zabs = 2.139 toward Tol~1037$-$2703. The latter system
also shows detectable C~{\sc i$^*$} and C~{\sc ii$^*$} absorption
lines spread over $\sim70$ \kms and distributed in two distinct
velocity components without detectable H$_2$ absorption. 
Interestingly the only other system that shows C~{\sc i} 
absorption in the large programme data  is also found towards PKS~0237$-$2703.
This line of sight is famous for the presence of a super-cluster of C~{\sc iv} 
absorption lines. Thus, our study provides yet another motivation to
go for further deep imaging and follow-up spectroscopic observations
of the field (see Aragon-Salamanca et al. 1994).

%
%%%%%%%%%%%%%%%%%%%%%%%%%%%%%

\section*{acknowledgements} RS and PPJ gratefully acknowledge support
from the Indo-French
Centre for the Promotion of Advanced Research (Centre Franco-Indien pour
la Promotion de la Recherche Avanc\'ee) under contract No. 3004-3.
We thank GMRT staff for their co-operation during our observations.
The GMRT is an international facility run by National Centre for Radio
Astrophysics of Tata Institute of Fundamental Research.
Some of the data presented in this paper were obtained from the 
Multimission Archive at the Space Telescope Science Institute (MAST). 
STScI is operated by the Association of Universities for Research in 
Astronomy, Inc., under NASA contract NAS5-26555. Support for MAST for 
non-HST data is provided by the NASA Office of Space Science via grant 
NAG5-7584 and by other grants and contracts.
%
%
%%%%%%%%%%%%%%%%%%%%%%%%%%%%%


\begin{thebibliography}{}
%
\bibitem{} Abbott, D. C. 1982, ApJ, 263, 723
%
\bibitem{} Aracil, B., Petitjean, P.,  Pichon, C. \& Bergeron, J. 2004, A\&A, 419, 811
%
\bibitem{} Aragon-Salamanca, A., Ellis, R.S., Schwarzenberg, J.-M., Bergeron, J., 1994, ApJ, 421, 27
%
\bibitem{} Ballester, P., Dorigo, K., Disar\'o, A., Pizarro de La Iglesia, J. A., Modigliani, A., \& Boitquin, O., ASPC, 216, 461
%
\bibitem{} Bahcall, J., Joss P. C., Lynds, R., 1973, ApJ, 182, 95
%
\bibitem{} Bergeron, J., Petitjean, P., Aracil, B. et al. 2004, Msngr, 118, 40
%
\bibitem{} Black, J. H. 1987, Interstellar processes, ed. Hollenbach, D. J., Thronson, H. A. (Dordrecht: Reidel), p731
%
%\bibitem{} Boiss\'e, P., \& Bergeron, J. 1985, A\&A, 145, 59
%	 
\bibitem{} Briggs, F. H., \& Wolfe, A. M., 1983, ApJ, 268, 76	
%
\bibitem{} Briggs, F. H., Wolfe, A. M., Liszt, H. S. et al. 1989, ApJ, 341, 650 
%
\bibitem{} Carilli, C., Land, W., de Bruyn, A. G., Braun, R., \&
Miley, G. K. 1996, AJ, 112, 1317
%
\bibitem{} Chand, H., Srianand, R., Petitjean, P., \& Aracil, B. 2004, A\&A,
417, 853
%
\bibitem{} Curran, S. J., Murphy, M. T., Philstrom, Y. M., Webb, J. K., \&
Purcell, C. R. 2005, MNRAS,356, 1509
%
\bibitem{} Dekker H., D'Odorico S., Kaufer A., Delabre B., \& Kotzlowski H.
           2000, in Iye M., Moorwood A. F., eds, Proc. SPIE Vol. 4008,
           Optical and IR telescope instrumentation and detectors, p. 534
%
\bibitem{} Draine, B. T., \& Bertoldi, F. 1996, ApJ, 468, 269
%
\bibitem{} Edl\'en, B. 1966, Metrologia, 2, 71
%
\bibitem{}  Ellison, S. L., Ibata, R., Pettini, M., Lewis, G. F., 
Aracil, B., Petitjean, P. \& Srianand, R. 2004, A\&A, 414, 79
%
\bibitem{} Ferland G. J., Korista, K. T., Verner, D. A., Ferguson, J. W., Kingdon, J. B., \& Verner, E. M. 1998, PASP, 110, 761 
%
\bibitem{} Fey, A. L., Clegg, A. W., \& Fomalont, E. B. 1996, ApJS, 105, 299
%
\bibitem{} Fomalont, E. B., Frey, S., Paragi, Z. et al., 2000, ApJS, 131, 95
%
\bibitem{} Ge, J., \& Bechtold, J. 1999, in Carilli C. L., Radford S. J. E.,
Menten K. M., \& Langston G. I., eds., Highly redshifted Radio Lines,
ASP Conf. Series, Vol. 156, p. 121

%
\bibitem{} Ge, J., Bechtold, J. \& Black, J. H. A. 1997, ApJ, 434, 67
%
\bibitem{} Haardt, F  \& Madau, P., 2005, Private communication.
%
\bibitem{} Heinm\"uller, J., Patrick, P., Ledoux, C., Caucci, S., \& Srianand, R. 
2005, A\&A, 449, 33
%
\bibitem{} Jenkins, E. B., \& Tripp, T. M. 2001, ApJ, 137, 297
%
\bibitem{} Kanekar, N., \& Chengalur, J., 2003, A\&A, 399, 857
%
\bibitem{} Lane W., 2000, PhD Thesis, University of Groningen 
(http://www.astron.nl/p/WSRT\_thesis.htm)
%
%\bibitem{} Lanzetta, K. M., Wolfe, A. M., \& Turnshek, D. A. 1995, ApJ, 440, 435
%
\bibitem{} Ledoux, C., Petitjean, P., \& Srianand, R. 2003, MNRAS, 346,209
%
\bibitem{} Ledoux, C., Srianand, R., \& Petitjean, P., 2002, A\&A, 392, 781
%
\bibitem{} McKee. C. F.,  \& Ostriker, J. P. 1977,ApJ, 218, 148
%
\bibitem{} Meyer, D. M., Black, J. H., Chaffee, F. H., Foltz, G., \& York, D.,
1986, ApJ, 308, 37
%
\bibitem{} Petitjean, P., Aracil, B., Srianand, R. \& Ibata, R. 2000, A\&A, 359, 457
%
\bibitem{} Petitjean, P., Srianand, R., \& Ledoux, C. 2000, A\&A, 364, L26
%
\bibitem{} Petitjean, P., Srianand, R., \& Ledoux, C. 2002, MNRAS, 332, 383
%
\bibitem{} Rao, S. M., \& Turnshek, D. A. 2000, ApJS, 130, 1
%
\bibitem{} Richter, P., Ledoux, C., Petitjean, P. \& Bergeron, J.  2005, A\&A, 440, 819
%
\bibitem{} Roth, K. C., \& Bauer, J. 1999, ApJ, 515, 57
%
\bibitem{} Shaw, G., Ferland, G., Srianand, R., \& Abel, N. 2006, ApJ, 639, 941 
%
\bibitem{} Silva, A. I., \& Viegas, S. M. 2002, MNRAS, 329, 135
%
\bibitem{} Songaila, A et al., 1994, Nature, 371, 43
%
\bibitem{} Srianand, R., \& Petitjean, P. 1998, A\&A, 337, 51
%
\bibitem{} Srianand, R., \& Petitjean, P. 2001, A\&A, 373, 816
%
\bibitem{} Srianand, R., Petitjean, P., \& Ledoux, C., 2000, Nature, 408, 931
%
\bibitem{} Srianand, R., Petitjean, P., Ledoux, C., Ferland, G., \& Shaw, G. 2005,
MNRAS, 362, 549
%
\bibitem{} Stumpff, P. 1980, A\&AS, 41, 1
%
\bibitem[]{}Viegas, S. M. 1995, MNRAS, 272, 35
\bibitem[]{} Wolfe, A. M., \& Briggs, F.H. 1981, ApJ, 248, 460
\bibitem[]{} Wolfe, A. M., Briggs, F. H., Turnshek, D. A., Davis, M. M., Smith, H. E., \& Cohen, R. D. 1985, ApJ, 294, L67 
\bibitem[]{}
Wolfe, A. M. Prochaska, J. X., \& Gawiser, E. 2003, ApJ, 593, 215
\bibitem[]{}
Wolfe, A. M., Gawiser, E., \& Prochaska, J. X. 2003, ApJ, 593, 235
\bibitem[]{}
Wolfe, A. M., Howk, J. C., Gawiser, E., Prochaska, J. X., \& Lopez, S. 2004, ApJ, 615, 525
\bibitem{} Yamasaki, N. Y., Miyazaki, H., Ohashi, T., \& Wilkes, B. 1998, PASJ, 50, 19

\end{thebibliography}
\end{document}